\documentclass[aps,prl,twocolumn,groupedaddress,showpacs,floatfix]{revtex4}

\usepackage{bm}
\usepackage{amssymb}
\usepackage{amsmath}
\usepackage[dvips]{graphicx}

\begin{document}


\title{Non mean-field behaviour of critical wetting transition for short-range forces.}


\author{Pawe{\l}  Bryk}
\affiliation{Department for the Modeling of Physico--Chemical Processes, Maria Curie--Sk{\l}odowska University, 20--031
Lublin, Poland.}
\email[]{pawel.bryk@gmail.com}
\author{Kurt Binder}
\affiliation{Institu f{\"u}r Physik, Johannes Gutenberg-Universit{\"a}t Mainz, Staudinger Weg 7, D-55099 Mainz, Germany}


\newcommand{\prom}[2]{\left \langle #1 \right \rangle_{#2}}
\newcommand{\du}{\mbox{$\,$} \mathrm{d}}
\newcommand{\Ppar}{p_{\parallel}}
\newcommand{\Pper}{p_{\perp}}
\newcommand{\derpar}[3]{
        \left(\frac{\partial #1}{\partial #2}\right)_{#3}
                      }
\newcommand{\Eq}[1]{Eq.~(\ref{#1})}
\newcommand{\rvec}[2]{\mathbf{r}_{#1}^{#2}}
\newcommand{\qvec}[2]{\mathbf{q}_{#1}^{#2}}
\newcommand{\Rvec}[2]{\mathbf{R}_{#1}^{#2}}
\newcommand{\tvec}[2]{\mathbf{t}_{#1}^{#2}}
\newcommand{\internaldistribution}[1]{P_{#1}}
\newcommand{\rhopoint}{\hat{\rho}}
\newcommand{\field}{v_{ext}}


\date{\today}

\begin{abstract}
Critical wetting transition for short-range forces in three dimensions ($d=3$) is reinvestigated
by means of Monte Carlo simulation. Using anisotropic finite size scaling approach, as well as
approaches that do not rely on finite size scaling, we show that the critical wetting
transition shows clear deviation from mean-field behaviour. We estimate that the
effective critical exponent $\nu_{\|}^{\textrm{eff}}=1.76\pm 0.08$ for $J/kT=0.35$ and $\nu_{\|}^{\textrm{eff}}=1.85\pm 0.07$ for $J/kT=0.25$. These values are in accord with predictions
of Parry {\it et al.} [Phys. Rev. Lett. {\bf 100}, 136105 (2008)].
We also point out that the anisotropic finite size scaling approach in $d=3$ requires additional
modification in order to reach full consistency of simulational results.
\end{abstract}


\maketitle

Understanding interfacial properties of fluids is important for
many applications including adsorption in porous materials\cite{Gelb99}, nanofluidic devices\cite{Bocquet10} and design of superhydrophobic surfaces\cite{Li07}.
The common problem pertinent to these research areas is the prediction
and control of the wetting properties of surfaces.
The introduction of patterns on the nanoscale leads to substantial changes
in wettability and creates a host of new effects\cite{Gang05,Hofmann10,Parry11}.
While the wetting phenomena at planar surfaces is well understood\cite{Bonn09}, there is one notable exception. Critical wetting has been a long standing and
stubbornly difficult problem to understand.

Renormalization group (RG) calulations based on a local interfacial Hamiltonian
\cite{Brezin83,Fisher85} predict that the critical wetting transition
for short range forces is strongly nonuniversal. 
When temperature $T$ approaches the wetting temperature $T_w$,
the critical exponent characterizing the divergence
of the paralell correlation length, $\xi_{\|}\sim (T_w - T)^{-\nu_{\|}}$, depends on
a nonuniversal dimensionaless wetting parameter, $\omega= \frac{k T}{4\pi\Sigma\xi^2}$,
where $k$ is Boltzmann's constant, $\Sigma$ is the interfacial stiffness (or surface tension for simple liquids),
and $\xi$ is the correlation length in the phase that wets the wall. For the case of
the Ising model in three dimensions ($d=3$) one has $1/2 < \omega < 2$, which leads to
\begin{equation}
\nu_{\|}(\omega)=(\sqrt{2}-\sqrt{\omega})^{-2}\;.
\end{equation}
Surprisingly, subsequent simulation studies \cite{Binder86,Binder88,Binder89}
showed only minor deviations from mean-field value, $\nu_{\|}^{\textrm{MF}}=1$.
In order to reconcile theory and simulation Parry {\it et al.}\cite{Parry08} proposed a new
non-local (NL) interface Hamiltionian which removed various intrinsic inconsistencies
of previous approaches. Reanalysis of the NL model showed the apperance of another
diverging length $\xi_{\textrm{NL}}=\sqrt{l\xi}\propto \sqrt{\ln\xi_{\|}}$, which cuts some
of the interfacial fluctuations for small film thicknesses $l$. This in turn leads
to a reduction in the effective value of the wetting parameter $\omega_{\textrm{eff}}$,
and to an effective exponent $\nu_{\|}^{\textrm{eff}}$.

\begin{figure}[tbh]
\includegraphics[width=8cm,clip]{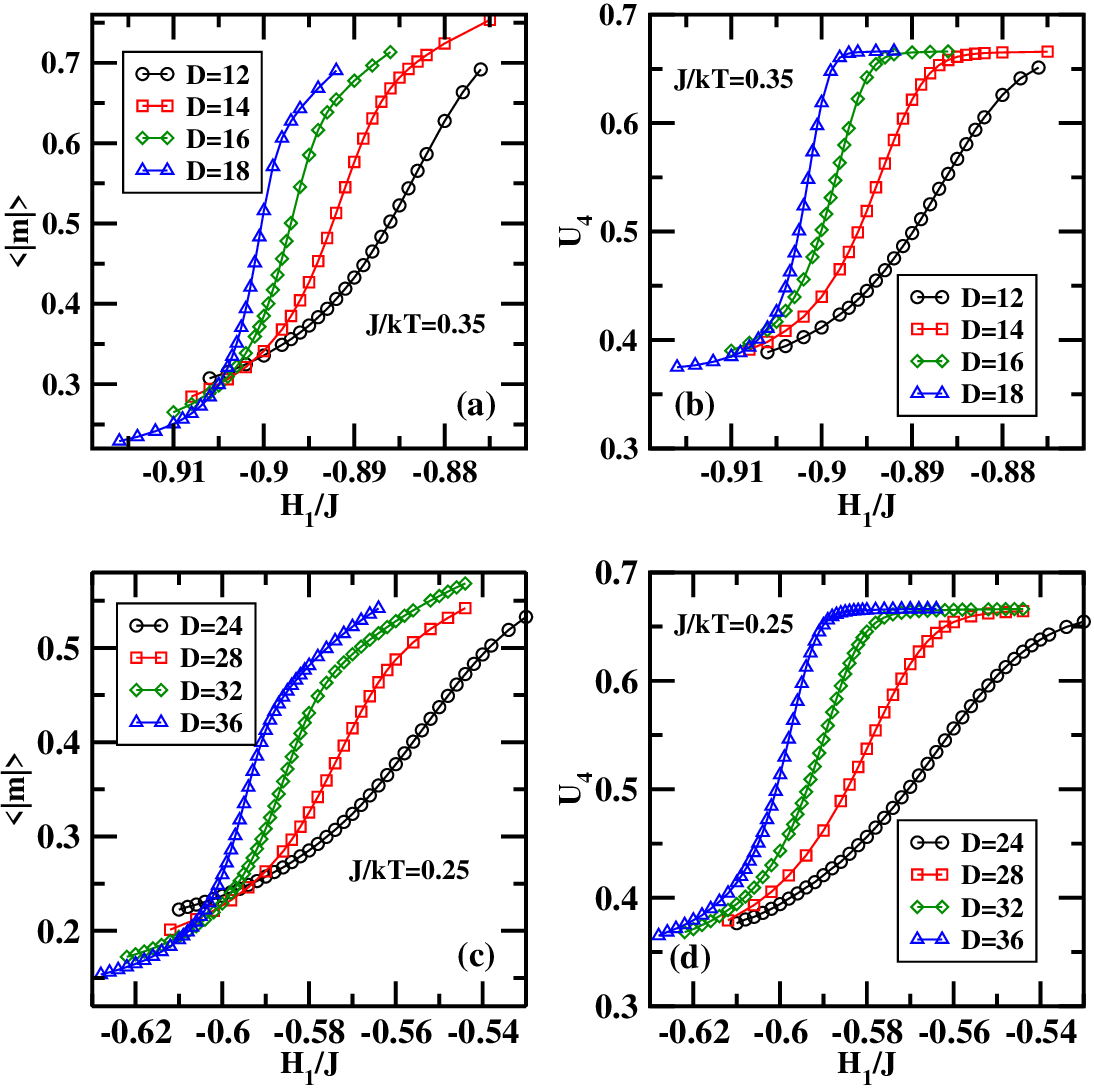}
\caption{\label{fig:1}
(color online) Average absolute magnetization $\langle |m| \rangle$, and cumulants $U_4$, vs.
surface field $H_1/J$. Parts (a) and (b) show the results evaluated at $J/k T=0.35$ and
for the generalized aspect ratio $C^*=2.8854$, while parts (c) and (d) show the results calculated
for $J/k T=0.25$ and for $C^*=5.77078$. System sizes $D$ are given in the Figure.
}
\end{figure}
Very recently a new {\it anisotropic} finite size scaling (AFSS) theory which should be suitable for studying wetting transitions in general was proposed \cite{Albano12a,Albano12b}. These authors have suggested that the previous \cite{Binder86,Binder88,Binder89} estimates for the location
of the critical wetting transition need a revision, however the critical exponent 
has not been determined. In this Communication we reconsider this approach and show that the critical wetting transition for short-range forces in $d=3$ shows clear deviations from mean-field theory. We also give evidence
that the AFSS theory is problematic in $d=3$, which was not anticipated before \cite{Albano12a,Albano12b} 

We consider simple-cubic Ising $L\times L\times D$ systems with two free surface layers $L\times L$,
and periodic boundary conditions in two remaining directions. The local
order parameter of the corresponding phase transition is a pseudospin variable
$s_i=\pm 1$ at lattice site $i$. The Hamiltonian for the system is
\begin{equation}
{\cal H}=-J\sum_{bulk}s_i s_j
-H \sum_{bulk}s_i -H_1 \sum_{k \in surf 1}s_k -H_D \sum_{k \in surf D}s_k\;,
\end{equation}
where $J$ is the bulk exchange constant, $H$ is the bulk field.
Surface fields $H_1$ and $H_D$ act only on the first and last layer, respectively.
In order to avoid effects connected to capillary condensation we select "antisymmetric"
walls, i.e. $H_1=-H_D<0$. 
During the course of simulation several quantities were accumulated, including
the average absolute value $\langle |m| \rangle$ of the magnetization
$m=(L^2D)^{-1}\sum_{i}s_i$,
susceptibility
$\chi_{'}=L^2 D (\langle m^2 \rangle - \langle |m| \rangle^2)/kT$, and the fourth order cumulant
$U_4=1- \langle m^4 \rangle /3 \langle m^2 \rangle^2$.
When the system is in the partial wetting
regime, the interface is bound to the wall $k=1$ or $k=D$ with equal probability.
Consequently, $\langle |m| \rangle$ is nonzero in the thermodynamic limit. On the other hand, for the wet state
the interface is unbound from either of the walls and wanders around the middle of the system.
Consequently, $\langle |m| \rangle$ is zero for $D\rightarrow \infty$.
The systems were simulated using highly efficient multispin coding algorithm \cite{Wansleben87}.
In order to overcome critical slowing down near the critical wetting point
we applied hyperparallel tempering technique \cite{Yan99} and simulated many systems
at the same time, and allowed for frequent swaps between them.
 Statistical effort was at least $5\times 10^7$ spin flips per site.
\begin{figure}[tbh]
\includegraphics[width=8cm,clip]{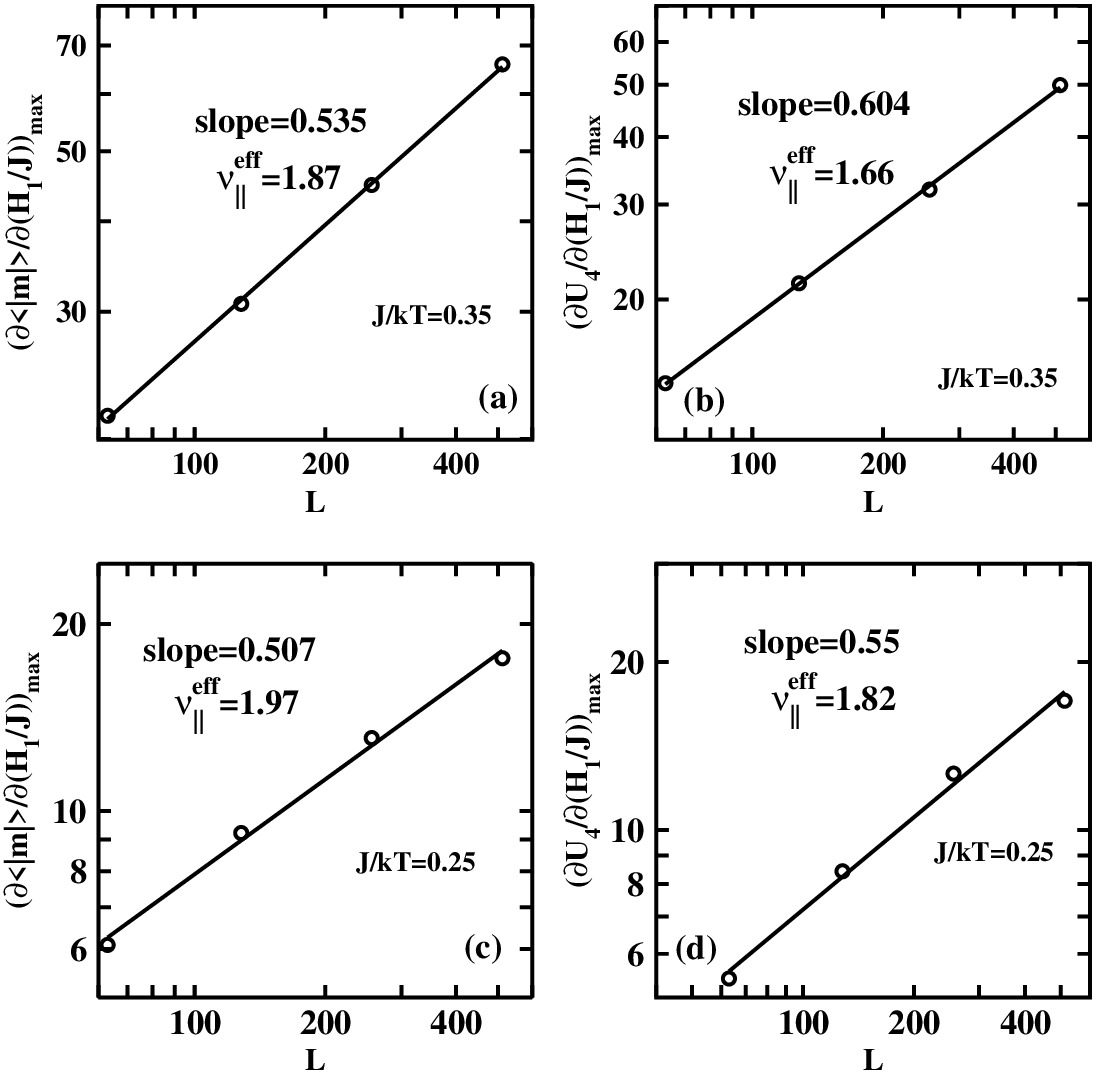}
\caption{\label{fig:2}
Estimation of the effective critical exponent $\nu_{\|}^{\textrm{eff}}$ using AFSS approach.
Plots show the maximum slope of the average absolute magnetization 
$(\partial \langle |m| \rangle/ \partial (H_1/J))_{max}$ and the maximum slope of
the cumulant $(\partial U_4/ \partial (H_1/J))_{max}$ vs. linear system size $L$.
Parts (a) and and (b) denote the results obtained for $J/kT=0.35$ while
parts (c) and (d) are for $J/kT=0.25$.
}
\end{figure}

Within AFSS approach the thermodynamic limit $D\rightarrow \infty$ must be taken
in a special way, keeping the generalized aspect ratio $C=D^{\nu_{\|}/\nu_{\perp}}/L$ (or, alternatively $C^*=D/L^{\nu_{\perp}/\nu_{\|}}$) constant\cite{Albano12a,Albano12b}.
The scaling ansatz for the order parameter probability distribution is given by
\begin{equation}
P_{D,L}(m)=\xi_{\|}^{\beta/\nu_{\|}}\tilde{P}(C,L/\xi_{\|},m\xi^{\beta/\nu_{\|}}_{\|}),\; m\rightarrow 0,\;\; \xi_{\|}\rightarrow\infty\;,
\end{equation}
where $\tilde{P}$ is a scaling function, whereas $\beta$ is the order parameter critical exponent. For $d=3$ $\beta=0$ while the exponent for the transverse correlation length $\nu_{\perp}=0$. Consequently we keep fixed the generalized aspect ratio of the form $C^*=D/\ln (L)$.

Following earlier papers \cite{Albano12a,Binder86,Binder88,Binder89} we keep the temperature constant (which keeps fixed the bulk correlation length) and vary the surface field $H_1$. The calculations were carried out for two temperatures, $J/kT=0.35$ with $C^*=2.8854$, and for $J/kT=0.25$ with $C^*=5.77078$, and for several lateral system sizes.

Figure \ref{fig:1} shows the plots of the average absolute magnetization and the cumulant vs. $H_1$ calculated for the two temperatures.
Unlike the case of $d=2$, where both $\langle|m|\rangle$ and $U_4$ exhibit rather well-defined
unique intersection points, here the cumulants hardly intersect and the intersections
of $\langle|m|\rangle$ have not converged to a unique location either.
The nonexistence of intersection points is not as serious a problem, as it looks at first sight. Finite size can cause a shift as well as a rounding of a transition. Both should scale in the same way, should a straightforward application of finite size scaling work.
However even then it is possible that the amplitude prefactor for the shift is much larger than the rounding. In such case one would find no intersections for the cumulant.

The large statistical effort together with hyperparallel tempering technique yielded
accurate, smooth data allowing for an estimation
of the exponent $\nu_{\|}^{\textrm{eff}}$.  It has been established \cite{Albano12a,Albano12b} that 
$(\partial \langle |m| \rangle/ \partial (H_1/J))_{max} \propto L^{1/\nu_{\|}^{\textrm{eff}}}$, giving a very convenient
way of determination of the critical exponent $\nu_{\|}^{\textrm{eff}}$.
Likewise, a similar relation holds for cumulants, $(\partial U_4/ \partial (H_1/J))_{max} \propto L^{1/\nu_{\|}^{\textrm{eff}}}$.
Figure \ref{fig:2} shows the plots of the maximum slopes of $\langle |m| \rangle$ and $U_4$ vs. $L$.
We find that for both temperatures $\nu_{\|}^{\textrm{eff}}$ is clearly different from the mean-field value $\nu_{\|}^{\textrm{MF}}=1$, and attains values slightly below 2.

\begin{figure}[tbh]
\includegraphics[width=8cm,clip]{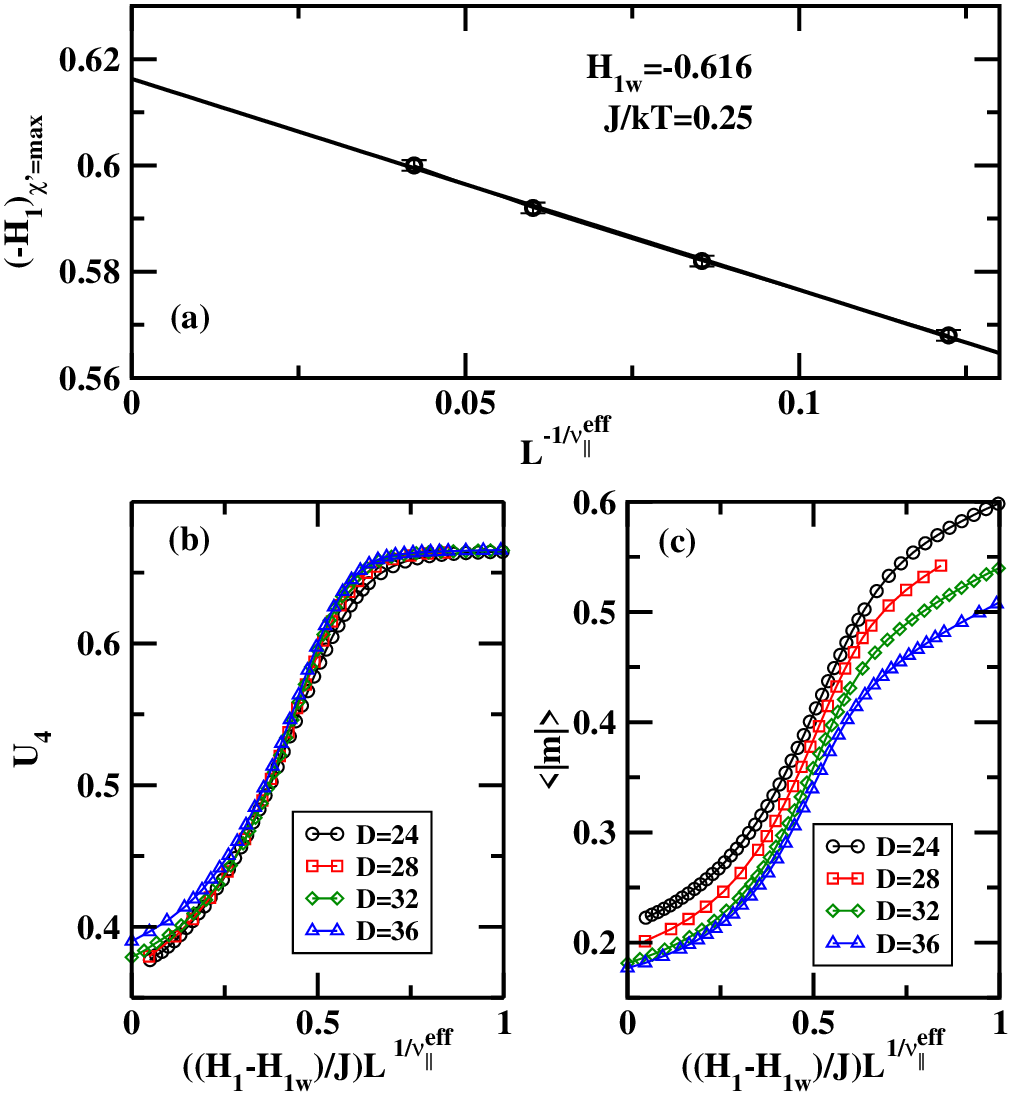}
\caption{\label{fig:3} 
(color online) (a) Estimation of the critical surface field for the wetting transition, $H_{1w}$,
for $J/kT=0.25$. The plot shows log-log plot of the position of the maximum susceptibility $\chi'$
vs $L^{-1/\nu_{\|}^{\textrm{eff}}}$. The intercept with $L^{-1/\nu_{\|}^{\textrm{eff}}}=0$ yields $H_{1w}$. (b) 
Scaling plot of $U_4$ vs $(H_1-H_{1w})L^{1/\nu_{\|}^{\textrm{eff}}}$ obtained using $H_{1w}=-0.616$ and $\nu_{\|}^{\textrm{eff}}=1.97$.
(c) Scaling plot of $\langle |m| \rangle$ vs $(H_1-H_{1w})L^{1/\nu_{\|}^{\textrm{eff}}}$ obtained using $H_{1w}=-0.616$ and $\nu_{\|}^{\textrm{eff}}=1.97$.
}
\end{figure}

Fig.~\ref{fig:3}(a) shows log-log plot of the position of maximum susceptibility $\chi'$ vs $L^{-1\nu_{\|}^{\textrm{eff}}}$ for $J/kT=0.25$.
We find that the value of the surface field for the wetting transition $H_{1w}=-0.616\pm 0.002$. 
Such a very good fit would not be possible if the mean-field value $\nu_{\|}^{\textrm{MF}}=1$ was used instead.
Further consistency checks are displayed in Fig.~\ref{fig:3}(b) and (c).
We find a good scaling of the cumulants [cf. Fig.~\ref{fig:3}(b)] with the estimated value of the effective exponent. 
Quite surprisingly, the plot of $\langle |m| \rangle$ vs. $(H_1-H_{1w})L^{1/\nu_{\|}^{\textrm{eff}}}$ do {\it not} collapse [cf. Fig.~\ref{fig:3}(c)].
Similar results were found for $J/kT=0.35$ \cite{supp}.
It seems, that there exists additional finite-size effect that should be applied to the ordinate variable $\langle |m| \rangle$!
These effects do not exist in $d=2$. The comparison of the diverging length scales indicates that for $d=3$  there is still one more divergence, $l_{eq}/\xi_{\perp} \propto \sqrt{\ln{\tau}}$ \cite{Dietrich88}, where $l_{eq}$ is the equilibrium film thickness, and $\tau$ is the distance from the transition. In contrast, for $d=2$ this ratio is constant. It is tempting to speculate, that the fact that the cumulants {\it do} collapse is connected with the fact, that these additional finite-size effect cancel out,
since $U_4$ is a ratio of moments of magnetization. Unfortunately, at present we do not see a straightforward way of incorporating this effect into AFSS framework.

\begin{figure}[tbh]
\includegraphics[width=8cm,clip]{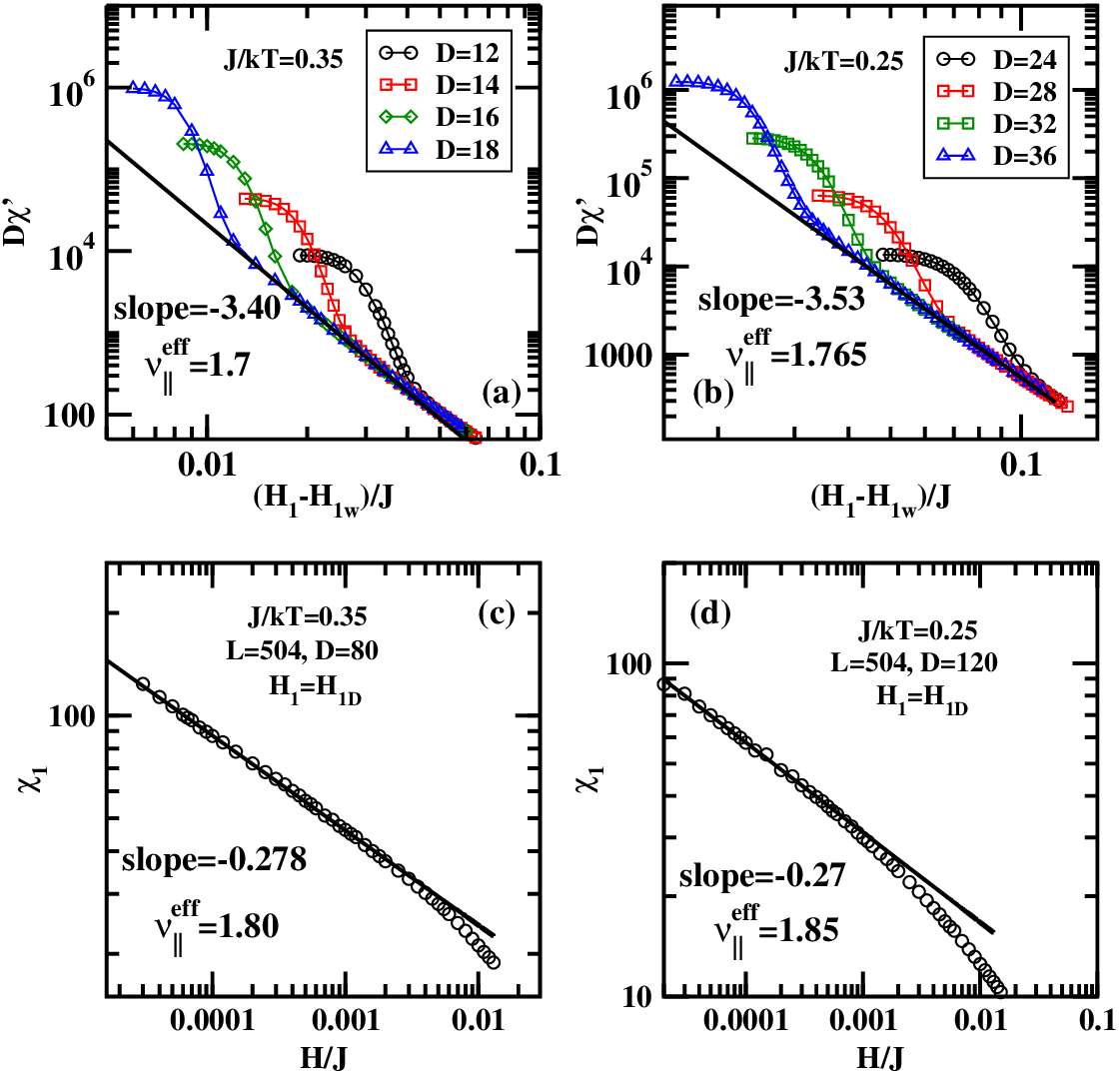}
\caption{\label{fig:4}
(color online) (a) and (b) Plots of the "mixed surface susceptibility"
vs $(H_1-H_{1w})J$.
(c) and (d) Plots of $\chi_s$ vs bulk field $H$ calculated for $H_1=H_{1w}$. The results were obtained using symmetric system, $H_1=H_D$ and for system sizes
given in the figure.}
\end{figure}

In view of the above it is natural to seek another evidence of the
non mean-field behaviour of critical wetting in $d=3$, which would not resort
to AFSS approach. It has been demonstrated \cite{Binder86,Binder88} that the "surface layer susceptibility"
$\chi_s=D\chi` \propto \xi_{\|}^2$. It follows, that when plotting 
$\chi_s$ vs. $H_1-H_{1w}$ for several system sizes, the regions unaffected
by finite size effects should exhibit the same slope equal to $2\nu_{\|}^{\textrm{eff}}$.
This provides additional estimation of the paralell correlation length exponent, independent of finite size scaling.
Figs.~\ref{fig:4}a and b demonstrate that for both temperatures the slope of $\chi_s$ in the region free of finite size effects
is a bit less than 4, which is consistent with previous estimates for $\nu_{\|}^{\textrm{eff}}$.

As a final check, in Figs.~\ref{fig:4}(c)-(d) we show log-log plots of the surface
susceptibility vs. non-zero bulk field $H$ evaluated at the critical surface fields $H_{1w}$,
for the two temperatures in question. The calculations presented here were carried out using
the "symmetric" boundary conditions, i.e. $H_1=H_D$ in order to follow exactly the computational
procedure presented in the first simulational studies on critical wetting \cite{Binder86,Binder88,Binder89}.
In such a system wetting films develop independently on both walls. 
During the simulations we monitor "surface layer susceptibility",
$\chi_{s}=\partial m_1 /\partial H=L^2 D (\langle m_1 m\rangle-\langle m_1 \rangle \langle m \rangle)$.
Since $\chi_s \sim H^{-1/2\nu_{\|}^{\textrm{eff}}}$ for $H>0$ at $H_{1}=H_{1w}$, the slope gives information about
the universality class of the wetting transition\cite{Binder88}.
The mean-field behavior would imply a slope of -0.5 and such was the conclusion of the early reports.
However, when the calculations are performed for the new estimations of the critical surface fields $H_{1w}$,
we observe clear deviations from mean-field exponents, again consistent with values obtained by different methods.
The final values of $\nu_{\|}^{\textrm{eff}}$ are obtained by averaging the exponents resulting from four different
methods. Putting together all the results we estimate that $\nu_{\|}^{\textrm{eff}}=1.76\pm 0.08$ for $J/kT=0.35$ and $\nu_{\|}^{\textrm{eff}}=1.85\pm 0.07$ for $J/kT=0.25$. Once the exponents are determined, we are able to calculate
the effective wetting parameter. We obtain $\omega_{\textrm{eff}}=0.44\pm 0.03$ and $0.46\pm 0.02$ for $J/kT=0.35$ and
0.25, respectively. 

Our results indicate that the early conclusions about the mean-field behaviour of critical wetting in $d=3$ 
can be traced back to the inaccurate estimation of the critical surface field $H_{1w}$. 
This is not to say that those simulations were wrong. Simply, using the computing resources avaiable almost thirty
years ago it was not possible to arrive at the correct conclusions. 
It is now clear that sizes like $L=50$ \cite{Binder86,Binder88} were far too small.
The fact that even for the lateral system size $L=504$
we reach roughly only half of the full nonuniversal value of $\nu_{\|} \approx 3.7$ is in accordance with theoretical conjecture
of Parry {\it et al.}\cite{Parry08} about very slow crossover to the asymptotic regime. We estimate that the system sizes
required to see in simulations the full non-universal behaviour of critical wetting must be of the order of
tens of thousands lattice spacings.

In conclusion, we have carried out accurate Monte Carlo simulations of critical wetting transition in $d=3$. We have found clear deviations from mean-field behaviour. We estimate that the effective critical exponent
$\nu_{\|}^{\textrm{eff}}=1.71\pm 0.1$ for $J/kT=0.35$ and $\nu_{\|}^{\textrm{eff}}=1.76\pm 0.17$ for $J/kT=0.25$.
Our results clearly support the non-local Hamiltonian model \cite{Parry08} and together with Ref.\onlinecite{Pang11}
(where related effects due to $\xi_{NL}$ for complete wetting were studied)
provide a strong evidence towards the validity of this approach.
We have also found that the understanding of finite size effects on critical wetting in $d = 3$ is still incomplete: analytical guidance to find the proper extension of the AFSS approach to cope with the weak logarithmic divergence of
the perpendicular correlation length remains a future challenge, to reach a full understanding of the simulation results. Thus, a longstanding puzzle may finally be close to its resolution.

\begin{figure}[tbh]
\includegraphics[width=8cm,clip]{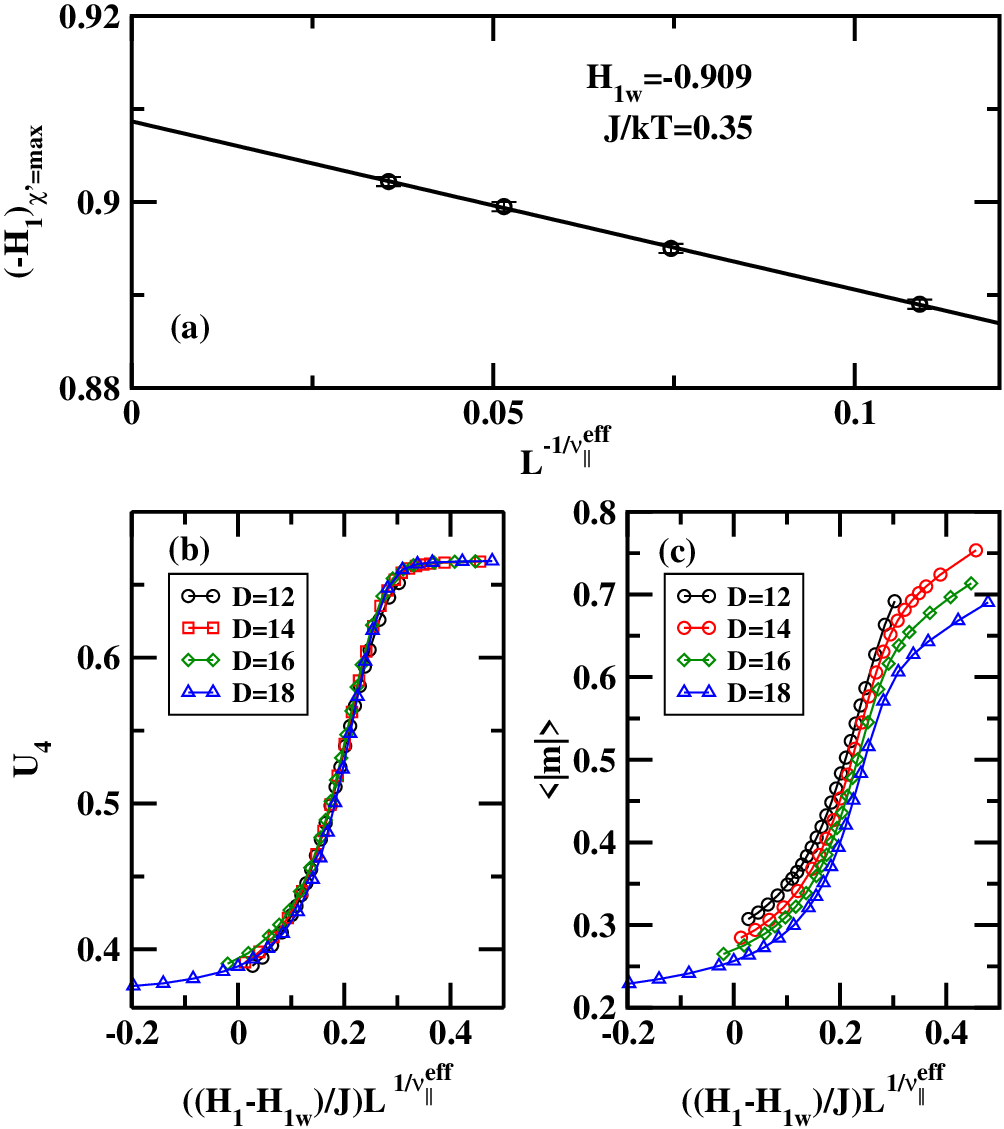}
\caption{\label{fig:Supplemental material} 
(color online) Supplemental material: (a) Estimation of the critical surface field for the wetting transition, $H_{1w}$,
for $J/kT=0.35$. The plot shows log-log plot of the position of the maximum susceptibility $\chi'$
vs $L^{-1/\nu_{\|}^{\textrm{eff}}}$. The intercept with $L^{-1/\nu_{\|}^{\textrm{eff}}}=0$ yields $H_{1w}$. (b) 
Scaling plot of $U_4$ vs $(H_1-H_{1w})L^{1/\nu_{\|}^{\textrm{eff}}}$.
(c) Scaling plot of $\langle |m| \rangle$ vs $(H_1-H_{1w})L^{1/\nu_{\|}^{\textrm{eff}}}$.
}
\end{figure} 

\begin{acknowledgments}
P.B is grateful to W. R{\.z}ysko and N.R. Bernandino for many discussions.
K.B. thanks E.V. Albano and D.P. Landau for discussions.
\end{acknowledgments}





\end{document}